\input jytex.tex   
\typesize=10pt
\magnification=1200
\baselineskip17truept
\footnotenumstyle{arabic}
\hsize=6truein\vsize=8.5truein
\sectionnumstyle{blank}
\chapternumstyle{blank}
\chapternum=1
\sectionnum=1
\pagenum=0

\def\begintitle{\pagenumstyle{blank}\parindent=0pt\begin{narrow}[0.4in]}
\def\endtitle{\end{narrow}\newpage\pagenumstyle{arabic}}


\def\beginexercise{\vskip 20truept\parindent=0pt\begin{narrow}[10
truept]}
\def\endexercise{\vskip 10truept\end{narrow}}


\def\eql#1{\eqno\eqnlabel{#1}}
\def\ref{\reference}
\def\peq{\puteqn}
\def\pref{\putref}

\def\mgn{\marginnote}
\def\bex{\begin{exercise}}
\def\eex{\end{exercise}}


\font\open=msbm10 
\font\goth=eufm10  

\def\mbox#1{{\leavevmode\hbox{#1}}}

\def\hspace#1{{\phantom{\mbox#1}}}
\def\oR{\mbox{\open\char82}}

\def\gK{\mbox{{\goth\char75}}}

\def\al{\alpha}
\def\be{\beta}

\def\Ga{\Gamma}

\def\ep{\epsilon}

\def\la{\lambda}

\def\om{\omega}

\def\si{\sigma}

\def\th{\theta}

\def\ze{\zeta}

\def\De{\Delta}

\def\zs{{\rm zs }}
\def\zn{{\rm zn }}

\def\zf{$\zeta$--function}


\def\frac#1/#2{\leavevmode\kern.1em
\raise.5ex\hbox{\the\scriptfont0 #1}\kern-.1em/\kern-.15em
\lower.25ex\hbox{\the\scriptfont0 #2}}
\def\sfrac#1/#2{\leavevmode\kern.1em
\raise.5ex\hbox{\the\scriptscriptfont0 #1}\kern-.1em/\kern-.15em
\lower.25ex\hbox{\the\scriptscriptfont0 #2}}

\def\gtorder{\mathrel{\raise.3ex\hbox{$>$}\mkern-14mu
             \lower0.6ex\hbox{$\sim$}}}
\def\ltorder{\mathrel{\raise.3ex\hbox{$<$}\mkern-14mu
             \lower0.6ex\hbox{$\sim$}}}

\def\semidirprod{\rlap{\ss C}\raise1pt\hbox{$\mkern.75mu\times$}}
\def\for{\lower6pt\hbox{$\Big|$}}
\def\fish{\kern-.25em{\phantom{abcde}\over \phantom{abcde}}\kern-.25em}


\def\boxit#1{\vbox{\hrule\hbox{\vrule\kern3pt
        \vbox{\kern3pt#1\kern3pt}\kern3pt\vrule}\hrule}}
\def\dalemb#1#2{{\vbox{\hrule height .#2pt
        \hbox{\vrule width.#2pt height#1pt \kern#1pt
                \vrule width.#2pt}
        \hrule height.#2pt}}}

\def\frac#1#2{{{#1}\over{#2}}}

\def\comb#1#2{{\left(#1\atop#2\right)}}

\def\cosec{{\rm cosec\,}}
\def\etc{{\it etc. }}

\def\eg{{\it e.g. }}
\def\ie{{\it i.e. }}
\def\cf{{\it cf }}
\def\pa{\partial}


\def\ol {\overline}
\def\sech{{\rm sech\,}}
\def\cosech{{\rm cosech\,}}

\def\sumdasht#1#2{{\mathop{{\sum}'}_{#1}^{#2}}}

\def\3j#1#2#3#4#5#6{\left\lgroup\matrix{#1&#2&#3\cr#4&#5&#6\cr}
\right\rgroup}

\def\caE{{\cal E}}
\def\caD{{\cal D}}
\def\caF{{\cal F}}

\def\m?{\mgn{?}}

\def\pa{\partial}

\def\beq{\begin{eqnarray}}
\def\eeq{\end{eqnarray}}


\def\cqg#1#2#3{{\it Class. Quant. Grav.} {\bf {#1}} ({#2}) #3}

\def\jmp#1#2#3{{\it J. Math. Phys.} {\bf {#1}} ({#2}) #3}
\def\jpa#1#2#3{{\it J. Phys.} {\bf A{#1}} ({#2}) #3}

\def\np#1#2#3{{\it Nucl. Phys.} {\bf B{#1}} ({#2}) #3}
\def\pl#1#2#3{{\it Phys. Lett.} {\bf {#1}} ({#2}) #3}

\def\prp#1#2#3{{\it Phys. Rep.} {\bf {#1}} ({#2}) #3}
\def\pr#1#2#3{{\it Phys. Rev.} {\bf {#1}} ({#2}) #3}

\def\prD#1#2#3{{\it Phys. Rev.} {\bf D{#1}} ({#2}) #3}

\def\prs#1#2#3{{\it Proc. Roy. Soc.} {\bf A{#1}} ({#2}) #3}

\def\amsh#1#2#3{{\it Abh. Math. Sem. Ham.} {\bf {#1}} ({#2}) #3}
\def\am#1#2#3{{\it Acta Mathematica} {\bf {#1}} ({#2}) #3}
\def\aim#1#2#3{{\it Adv. in Math.} {\bf {#1}} ({#2}) #3}

\def\aom#1#2#3{{\it Ann. of Math.} {\bf {#1}} ({#2}) #3}

\def\dmj#1#2#3{{\it Duke Math. J.} {\bf {#1}} ({#2}) #3}
\def\invm#1#2#3{{\it Invent. Math.} {\bf {#1}} ({#2}) #3}

\def\jram#1#2#3{{\it J. f. reine u. Angew. Math.} {\bf {#1}} ({#2}) #3}
\def\jims#1#2#3{{\it J. Indian. Math. Soc.} {\bf {#1}} ({#2}) #3}
\def\jlms#1#2#3{{\it J. Lond. Math. Soc.} {\bf {#1}} ({#2}) #3}

\def\ma#1#2#3{{\it Math. Ann.} {\bf {#1}} ({#2}) #3}

\def\mz#1#2#3{{\it Math. Zeit.} {\bf {#1}} ({#2}) #3}

\def\plms#1#2#3{{\it Proc. Lond. Math. Soc.} {\bf {#1}} ({#2}) #3}
\def\pgma#1#2#3{{\it Proc. Glasgow Math. Ass.} {\bf {#1}} ({#2}) #3}
\def\qjm#1#2#3{{\it Quart. J. Math.} {\bf {#1}} ({#2}) #3}

\def\rmjm#1#2#3{{\it Rocky Mountain J. Math.} {\bf {#1}} ({#2}) #3}

\def\tams#1#2#3{{\it Trans.Am.Math.Soc.} {\bf {#1}} ({#2}) #3}

\vglue 1truein
\vskip15truept
\centertext{\Bigfonts\bf Elliptic aspects of statistical} \vskip 10truept
\centertext{\Bigfonts \bf mechanics on spheres}\vskip10truept
\centertext{\Bigfonts \bf }

 \vskip 20truept
\centertext{J.S.Dowker\footnote{ dowker@man.ac.uk} } \vskip
7truept \centertext{\it Theoretical Physics Group, }
\centertext{\it School of Physics and Astronomy} \centertext{\it
The University of Manchester,} \centertext{ \it Manchester,
England} \vskip 10truept \centertext{and} \vskip7truept
\centertext{Klaus Kirsten\footnote{ Klaus\_Kirsten@baylor.edu}}
\vskip7truept\centertext{\it Department of Mathematics, }
\centertext{\it Baylor University} \centertext{\it One Bear Place
\# 97328} \centertext{\it Waco, TX 76798 USA}
 \vskip15truept
 \begin{narrow}
Our earlier results on the temperature inversion properties and the
ellipticisation of the finite temperature internal energy on odd spheres
are extended to orbifold factors of odd spheres and then to other
thermodynamic quantities, in particular to the specific heat. The
behaviour under modular transformations is facilitated by the
introduction of a modular covariant derivative and it is shown that the
specific heat on any odd sphere can be expressed in terms of just three
functions. It is also shown that the free energy on the circle can be
written elliptically.

 \end{narrow}

\newpage
\section{\bf 1. Introduction.}

The topic of finite temperature free field theories on spheres has recently
reappeared specifically in connection with the Verlinde-Cardy relation. In
a previous paper, while examining this question, several technical points
arose which the present work intends to address. The equivalence of the
different ways of approaching the theory results in, or is a result of,
various mathematical identities, most of which have been around for some
time, many of them classic. It was thought useful to present this material
in a particular physical context. The opportunity will also be taken to
present some extensions of our previous programme.

In [\pref{DandK1}] two related aspects were considered. One was the
symmetry under temperature inversion and the other the `ellipticisation'
of the total internal energy $E$. It was shown, for example, that at
certain temperatures, $E$ could be expressed in finite terms involving
elliptic integrals.

In the present paper we discuss, firstly, the technical extension to
orbifold factors of spheres, S$^d/\Ga$. This analysis allows us to relate
several existing mathematical calculations.

All this detailed work concerns just the internal energy, $E$, and we
turn next to our major preoccupation which is a study of the other
thermodynamic quantities such as the free energy, entropy and specific
heat. Roughly speaking, these are related differentially to $E$.

Our objective, admittedly mathematical, is to determine the behaviour of
these quantities under modular transformations (which includes
temperature inversion) and to obtain elliptic expressions for them as far
as possible.

We find for the positive derivatives of the internal energy, such as the
specific heat, that the results generally parallel those of
[\pref{DandK1}] whereas for the negative derivatives (integrals), such as
the free energy, there are obstructions to temperature inversion symmetry
and ellipticisation. The nature of these obstructions makes contact with
some basic modular concepts, such period polynomials, which are not
pursued here.
\section{\bf 2. Resum\'e of earlier work.}

Our previous paper [\pref{DandK1}] was restricted to conformally invariant
free field theories on the space-times (`Einstein Universes')
$\oR\times$S$^d$ and, apart from a few comments, we exclusively discussed
odd spheres, S$^d$. We will continue to do so here.

The approach was based on individual treatment of the terms in the mode
degeneracy, which is just an odd polynomial in the mode label. The
complete sphere result was then obtained by simple combination of these
individual pieces. From this point of view there is nothing special about
the sphere. Any odd polynomial would do.

Other treatments, \eg [\pref{DandC,AandD}], yield more specific
forms. We begin here with an expression given by Chang and Dowker
[\pref{ChandD}] which is somewhat more general than just for the
sphere. The situation there discussed was a conformal scalar on an
orbifold factoring (`triangulation') of the sphere, written
S$^d/\Ga$, where $\Ga$ is a regular solid symmetry group.

The total free energy at temperature $1/\be$ on $\oR\times$S$^d/\Ga$ was
given in the form
  $$
  F=E_0-{1\over\be}\sum_{m=1}^\infty {1\over m} e^{\pm d_0m\be/2a}
\prod_{i=1}^d{1\over2\sinh(d_im\be/2a)}
  \eql{freeen}$$
where the $d_i$ are the degrees associated with the tiling group
$\Ga$ and $a$ is the radius of the sphere. $E_0$ is the zero
temperature, Casimir, value. It can be given in terms of Bernoulli
polynomials.

The hemisphere corresponds to $d_0=1$; there is only one reflecting
plane, and all the degrees $d_i$ are equal to 1. The upper sign in
(\peq{freeen}) refers to Neumann conditions on the rim of the hemisphere,
and the lower sign to Dirichlet conditions. To obtain the full sphere
periodic value these two expressions are added, and we will do this now
for {\it all} tilings so as to replace (\peq{freeen}) by
  $$
  F=E_0-{2\over\be}\sum_{m=1}^\infty {1\over m} \cosh(d_0m\be/2a)
\prod_{i=1}^d{1\over2\sinh(d_im\be/2a)}\,.
  \eql{freeen2}$$

For the time being we will work with just two examples; the hemisphere
(giving the periodic full sphere) and the quarter-sphere (giving the
periodic half-sphere). The quarter-sphere is a lune of angle $\pi/2$ and,
in this case, $d_0=2$, corresponding to the two reflecting planes, also
$d_1=2$, $d_i=1$ $(i=2,\ldots,d)$. There are conical singularities at the
north and south poles.

For these two examples, more specifically,
  $$
  F=E_0-{1\over2^{d-1}\be}\sum_{m=1}^\infty {1\over m} {\cosh(m\be/2a)
\over\sinh^d(m\be/2a)}\,,
  \eql{spfreeen}$$
and
  $$
  F=E_0-{1\over2^{d-1}\be}\sum_{m=1}^\infty {1\over m} {\coth(m\be/a)
\over\sinh^{d-1}(m\be/2a)}\,.
  \eql{4spfreeen}$$

Another way of writing (\peq{spfreeen}), which is sometimes useful, is,
  $$
   F=E_0+{a\over2^{d-2}(d-1)\be}{\pa\over\pa\be}\sum_{m=1}^\infty {1\over m^2}
   {1\over\sinh^{d-1}(m\be/2a)}\,.
  $$

As in our previous article, [\pref{DandK1}], we will work with the
total internal energy $\overline E = aE=a\pa(\be F/\pa\be)$,
  $$
 \overline E={\overline E}_0+{1\over2^{d}}\sum_{m=1}^\infty \bigg({d-1
\over\sinh^{d-1}(m\be/2a)}+{d\over\sinh^{d+1}(m\be/2a)}\bigg)\,,
  \eql{speen}$$
and
  $$
 {\overline E}={\overline E}_0+{1\over2^{d+1}}\sum_{m=1}^\infty \bigg({\sech^2(m\be/2a)
\over\sinh^{d+1}(m\be/2a)}+
{d-1\over\sinh^{d+1}(m\be/2a)}+{2(d-1)\over\sinh^{d-1}(m\be/2a)}\bigg)\,.
  \eql{4speen}$$

The full sphere expression (\peq{speen}) involves just the summation
  $$
   Q_{g}(x)=\sum_{m=1}^\infty \cosech^{2g}m x
  \eql{sum1}$$
and we now show how to reduce (\peq{4speen}) employing also the sum,
  $$
  R_{f}(x)=\sum_{m=1}^\infty \sech^{2f}m x\,.
  \eql{are}$$

Starting from the more general sum
  $$
  S_{g,f}(x)=\sum_{m=1}^\infty\sech^{2f}m x\,\,\cosech^{2g}m x
  \eql{sc}$$
and writing either
  $$
  \sech^2m x=1-\sinh^2m x\,\,\sech^2m x
  \eql{s}$$
or
  $$
  \cosech^2m x=\cosh^2m x\,\,\cosech^2m x-1
  \eql{c}$$
we find, for example from (\peq{s}), the recursion
  $$
  S_{g,f}=\sum_{j=0}^f(-1)^j\comb fj S_{g-j,j}
  $$
which is easily implemented starting from $S_{g,0}=Q_g$ and $S_{0,f}=R_f$.
In particular
  $$
  S_{g,1}=\sum_{j=1}^g (-1)^{g-j}\,Q_j+(-1)^g\,R_1
  $$
and the energies can be put in the form, if $d=2r+1$,
  $$
 {\overline E}_{2r+1}={\overline E}_0+{1\over2^{2r+1}}\big(2r\,Q_{r}+(2r+1)\,Q_{r+1}\big)
  \eql{en1}$$
and
  $$
   {\overline E}_{2r+1}={\overline E}_0+{1\over2^{2r+2}}\bigg((4r-1)\,Q_{r}+(2r+1)\,Q_{r+1}+\sum_{j=1}^{r-1}
   (-1)^{r+1-j}\,Q_j-(-1)^r\,R_1\bigg)
  \eql{en2}$$
where the arguments of all the $Q$'s and $R$'s are $\be/2a$.

In these two cases at least, we see that the problem reduces to the
calculation of the sums, (\peq{sum1}) and (\peq{are}), which are the
subjects of the papers by Kiyek and Schmidt [\pref{KandS}], Ling
[\pref{Ling}] and Zucker, [\pref{zucker2}]. Values of the sum, $Q_g$, for
specific $g$ are quite old.
\section{\bf 3. Hyperbolic summations.}
The methods in the papers just referred to, involve elliptic functions, in
one way or another, which is not surprising in view of the partial fraction
representation of hyperbolic functions and the double sum form of elliptic
functions which originated with Eisenstein, [\pref{Weil}].

Going back to the beginning, Eisenstein, following Euler, discussed the
single sums (see Hancock [\pref{Hancock}] p.32 Ex.5),
  $$
  (2g,x)=\sum_{n=-\infty}^\infty {1\over(x+n)^{2g}}=\pi^{2g}
  \sum_{k=1}^g (-1)^{k+g} A_{2g,2k}\,\cosec^{2k}\pi x\,.
  \eql{sum3}$$
Eisenstein says that the coefficients $A_{2g,2k}$ are simply related to
Bernoulli numbers and are given by a recursion derived from continued
differentiation.

Transcribing to the hyperbolic case,
  $$
  (2g,ix)=\sum_{n=-\infty}^\infty {1\over(ix+n)^{2g}}=(-1)^g \pi^{2g}
  \sum_{k=1}^g A_{2g,2k}\,\cosech^{2k}\pi x\,.
  \eql{sum4}$$

The explicit form of the recurrence is given by Ling, [\pref{Ling}], and no
doubt elsewhere after one hundred and sixty years. It is
  $$
  A_{2g+2,2k}={1\over2g(2g+1)}\big((2k-1)(2k-2)
   \,A_{2g,2k-2}+4k^2\,A_{2g,2k}\big)
  \eql{rec1}$$
and trivially $A_{2g,2g}=1$.

Now set $x=m\mu/\pi$ and sum again over $m$ referring to the definition
(\peq{sum1}). One finds,
  $$
  \sumdasht{m,n=-\infty}{\infty} {1\over(i\mu m/\pi+n)^{2g}}-2\ze_R(2g)
  =-2\pi^{2g}\sum_{k=1}^g A_{2g,2k}\,Q_k(\mu)\,,
  \eql{sum6}$$
and we have arrived at an {\it Eisenstein series}.

Equation (\peq{sum6}) can be used to determine $Q_k$ recursively
in terms of the left hand side which we have shown in
[\pref{DandK1}] is given in elliptic function terms. (This is a
standard result.) Such is Ling's method of finding the sums $Q_k$
and is a little roundabout. A further instalment is given in
[\pref{Ling2}].

The energy (\peq{en1}) can then be determined. Again one has to combine the
Eisenstein series values and the calculation is actually not so different
from our earlier one. The only novelty is that there is now no {\it
explicit} mention of the mode degeneracies. They are, however, lurking in
the analysis as the following shows.

An alternative way of arranging the evaluation of the $Q_g$ is to rewrite
the definition (\peq{sum1}) as a $q$-series as done by Zucker,
[\pref{zucker2}]. We re-express his approach somewhat to fit our
requirements. Resummation produces
  $$\eqalign{
  Q_g(\mu)&=2^{2g}\sum_{n=0}^\infty\comb{n+2g-1}n{q^{2(n+g)}\over1-q^{2(n+g)}}\cr
  &={2^{2g}\over(2g-1)!}\sum_{n=0}^\infty \big((n^2-(g-1)^2)\ldots n\big)
  {q^{2n}\over1-q^{2n}}\,,}
  \eql{cue}$$
where $q=e^{-\mu}$, and, not surprisingly, we see here the scalar
degeneracies making an effective appearance. The full sphere degeneracies
emerge on combining the two terms in (\peq{en1}) using (\peq{cue}).

Since they do not occur in our earlier work on the full sphere we now
discuss the $\sech$ sums, (\peq{are}) which follow from a translation of
the argument.

To begin with, quite easily from (\peq{sum3}) and (\peq{sum4}),
  $$
  (2g,{1\over2}-ix)=2^{2g}\sum_{n=-\infty}^\infty {1\over(i2x+2n+1)^{2g}}=\pi^{2g}
  \sum_{k=1}^g(-1)^{k+g} A_{2g,2k}\,\sech^{2k}\pi x\,,
  \eql{sum8}$$
and, corresponding to (\peq{sum6})
  $$
  2^{2g}\sum_{m,n=-\infty}^\infty {1\over(i 2m\mu/\pi+2n+1)^{2g}}-2
  \ze_R(2g,1/2)
  =2\pi^{2g}\sum_{k=1}^g(-1)^{k+g} A_{2g,2k}\,R_k(\mu)\,,
  \eql{sum7}$$
which contains an `even-odd' Eisenstein series. Reference to the series in
Glaisher [\pref{Glaisher}] shows that the Jacobi function zn is involved
this time. This has the expansion
  $$
  \gK\,\zn\, u=-\tan x-4\sum_{n=1}^\infty(-1)^{n-1}{q^{2n}\over1-q^{2n}}
  \sin 2nx\,,
  $$
where $x=\pi u/2K$ and $\gK =2K/\pi$, which can  also be obtained
from that for $\gK\,\zs\,u$ by adding $\pi/2$ to $x$. Here, as
usual, $q=e^{-\mu}=e^{-\pi K'/K}$, and $K,K'$ are given by $$K(k)
= \int\limits_0^1\frac{dz}{\sqrt{(1-z^2)(1-k^2z^2)}}, \quad \quad
K'=K (k'),\eql{K(k)}$$ where $k^2 + k'^2=1$.

Equivalently the $q$-series can be constructed, as for cosech,\mgn{DO}
  $$\eqalign{
  R_g(\mu)&=2^{2g}\sum_{n=0}^\infty (-1)^n
  \comb{n+2g-1}n{q^{2(n+g)}\over1-q^{2(n+g)}}\cr
  &=(-1)^{g}\,{2^{2g}\over(2g-1)!}\sum_{n=0}^\infty
  \big((n^2-(g-1)^2)\ldots n\big)
  (-1)^n{q^{2n}\over1-q^{2n}}\,,}
  \eql{cue2}$$
\cf Zucker, [\pref{zucker2}], who uses the expansion of nc$^2$ to extract the
Lambert series.

As has been mentioned, our previous analysis [\pref{DandK1}] expanded the
degeneracies in powers of $n^2$, $n$ being the mode label, and treated
each power separately making use of classic elliptic facts. This is the
same as Zucker's method and he, more or less, reproduces the degeneracy
expansion obtained, \eg by Cahn and Wolf, [\pref{CandW}], as well as the
elliptic expansions to be found in Glaisher, [\pref{Glaisher}]. Zucker's
general method of finding closed forms for the sums $Q_g(\mu)$ when $\mu$
corresponds to a singular modulus is thus entirely equivalent to our own
programme on the sphere, [\pref{DandK1}]. By the same token, the energy
on the periodic half--sphere, (\peq{4speen}), can also be expressed in
finite terms using the $S_{g,f}$, reduced to the $Q_g$ and $R_1$. It does
not seem possible to repeat this statement for the remaining tilings,
(\peq{freeen2}), at least not obviously.

The rather negative moral of this calculation is that, in order to
evaluate the summations that occur for the sphere expressions,
(\peq{freeen2}), in elliptic terms it seems necessary to re-introduce the
degeneracies in one way or another and to treat the individual terms, as
in our earlier work. The conclusion is that one might as well employ the
mode-degeneracy expressions in the first place, if these are known,
particularly since, in addition, some manipulation was required to obtain
(\peq{en1}) and (\peq{en2}).

Therefore, although the expressions (\peq{freeen2}) are particularly
compact for all the tilings, and provide adequately convergent numerical
forms, the earlier piecemeal approach has elliptic advantages and so we
turn to the degeneracy--mode approach for the tilings, knowing that an
elliptic formulation is possible for the quarter--sphere case at least.
We would anticipate that this should be true for all the odd sphere
`periodic' tilings because the heat--kernel expansion terminates, just as
it does on the full odd sphere. If this is so, then we could turn the
calculation around and produce exact forms for the `new' hyperbolic
summations in (\peq{freeen2}). This will not be attempted here.
\section{\bf 4. Mode forms applied to the quarter--sphere.}
In this section we detail the eigenvalue form of the thermodynamical
quantities and, as a typical example, consider the specific case of the
quarter-sphere, which has been treated previously [\pref{DandA}] in
another connection.

The Laplacian eigenvalues can be written in the form $\la=\om^2/a^2$
with, [\pref{ChandD}],
  $$
  \om_{m,n}=(\al+2n+m)\,,\quad n,m=0,1,\ldots
  $$
and the degeneracies are, ($d>1$),
  $$
  \comb{m+d-2}{d-2}\,.
  $$
The parameter $\al=(d+3)/2$ for Dirichlet (D) conditions, and $a=(d-1)/2$
for Neumann (N). The D and N expressions are added to get the `periodic'
(P) form.

For odd spheres we set $d=2r+1$. For $d\geq 5$ we shift the $m$ label by
$r+1$ in the D part and by $r-1$ in N, so that the eigenvalues and {\it
periodic} degeneracies, $g_{m,n}$, read, effectively,
  $$\eqalign{
  \om_{m,n}=&(2n+m+1)\,,\quad n=0,1,\ldots\,,\quad m=1,2,\ldots\cr
 g_{m,n}=&
  \comb{m+r-2}{2r-1}+\comb{m+r}{2r-1}\cr
  &\equiv g_m\,.}
  $$
We are allowed to extend and adjust the range of $m$ by the vanishing of
the binomials. For $d=3$ attention is needed because the label $m$ runs over
0,1,.... We continue with $d\ge 5$ and for $d=3$ we only have to add the
contribution from the $m=0$ mode at the end.

The degeneracies are made more explicit by remarking that $g_m$ is
polynomial in odd powers of $m$. Precisely,
  $$
  g_m= m\sum_{k=0}^{r-1} c_k m^{2k}\,.
  $$
These are of course related to Stirling polynomials, but we will leave them
as they are for the time being.

To proceed with the summation, residue classes mod 2 are introduced so that
$m=2l+p$ with $0\le l\le\infty$ and $p=0,1$, \ie $m$ is even (including 0)
or odd. The two $p$ values are treated separately at first. Then
  $$
  \om_{m,n}=2(n+l)+p+1=2N+p+1\,,
  $$
where a further advantageous relabelling has been made to the quadrant
coordinates,
  $$
   N=l+n\,,\quad{\rm and}\quad \nu=l-n\,,
  $$
so that $m=N+\nu+p$ with $N=0,1,\ldots$ and $-N\le\nu\le N$. Note that for
$N$ odd $\nu$ only takes odd values and for $N$ even $\nu$ only takes even
values.

We now write down the particular spectral quantity in which we are
interested; the internal energy,
  $$
  aE=\overline E=\overline E_0+\sum_{m,n =0}^{\infty}g_m\,\om_{m,n}\,
  {q^{2\om_{m,n}}\over
  1-q^{2\om_{m,n}}}\, ,
  $$
where the $m$ label is extended to 0 which is allowed because $g_0
=0$ and we introduced $q=e^{-\pi/\xi}$, $\xi = (2\pi a)/\beta$.
The sums are reorganised to allow for the $\nu$ independence of
the eigenvalues. Firstly
  $$
   \sum_{m,n=0}^{\infty}=\sum_{p=0,1}\,\sum_{l,n =0}^\infty\,,
  $$
and then
  $$
  \sum_{l,n =0}^\infty=\sum_{{N=0\atop even}}^\infty\,
  \sum_{{\nu=-N \atop  even}}^N + \sum_{{N=1 \atop  odd}}^\infty
   \sum_{{\nu = -N \atop  odd}} ^N \,.
  $$
This gives $$
  \overline E=\overline E_0+\sum_{p=0,1}\left(
  \sum_{{N=0\atop {\rm even}}}^\infty\,
  \sum_{{\nu=-N \atop even}}^N + \sum_{{N=1 \atop  odd}}^\infty
   \sum_{{\nu = -N \atop  odd}} ^N \right) g_{N+\nu +p}   (2N+p+1)
  \,{q^{2(2N+p+1)}\over
  1-q^{2(2N+p+1)}} .
  \eql{en7}
$$ The summation ranges, $N$ even and $N$ odd, are rewritten by $N\to 2N$
and $N\to 2N+1$ respectively. The relevant `degeneracies' are then
  $$\eqalign{
   g_{even}&=\sum_{\nu=-N}^N g_{2N+2\nu+p}=\sum_{k=0}^{r-1}\,c_k\,
\sum_{\nu=-N}^N\,(2N+p+2\nu)^{2k+1}\cr
   &=\sum_{k=0}^{r-1}c_k
2^{2k+1}\,\sum_{\nu'=p/2}^{2N+p/2}\nu'^{\,2k+1}\cr
&=\sum_{k=0}^{r-1}{c_k2^{2k}\over k+1}\,
\big(B_{2k+2}(2N+1+p/2)-B_{2k+2}(p/2)\big)}
  $$
and $$\eqalign{
  g_{odd} &=\sum_{\nu = -N-1}^N g_{2N+2\nu +p+2} =\sum_{k=0}^{r-1}
\,c_k\, \sum_{\nu = -N-1}^N (2N+p+2+2\nu)^{2k+1} \cr
   &=\sum_{k=0}^{r-1}c_k
2^{2k+1}\,\sum_{\nu'=p/2}^{2N+p/2+1}\nu'^{\,2k+1}\cr
 &=\sum_{k=0}^{r-1}{c_k2^{2k}\over k+1}\,
\big(B_{2k+2}(2N+2+p/2)-B_{2k+2}(p/2)\big) .}
  $$
Adding up all contributions, we have\mgn{Do vacuum}
  $$\eqalign{
  \overline E &= \overline E  _0 +
   \sum_{p=0,1}\sum_{N=0}^\infty \sum_{k=0}^{r-1}
   {c_k 2^{2k} \over k+1 } \times  \cr
   & \Big( \big( B_{2k+2}(2N+1+p/2)-B_{2k+2}(p/2)\big) (4N+p+1)
   {q^{2(4N+p+1)}\over 1-q^{2(4N+p+1)}} \cr
   & +\big( B_{2k+2}(2N+2+p/2)-B_{2k+2}(p/2)\big) (4N+p+3)
   {q^{2(4N+p+3)}\over 1-q^{2(4N+p+3)}} \Big)\,.}
  $$
Combining $p=0$ and $p=1$,
   $$
  \eqalign{
  \overline E &= \overline E  _0 +
   \sum_{k=0}^{r-1}
   {c_k 2^{2k} \over k+1 } \times  \cr
   & \sum_{n=1}^\infty \bigg( B_{2k+2}\big((n+1)/2\big)
   {n\,q^{2n}\over 1-q^{2n}}
   - B_{2k+2}(0)
   {(2n-1)\,q^{2(2n-1)}\over 1-q^{2(2n-1)}} \cr
   &\hspace{*******} - B_{2k+2} (1/2)  {2n\,q^{4n} \over 1-q^{4n}}\bigg) .}
  $$

In order to connect this to elliptic functions we want an explicit
expansion of the Bernoulli polynomials in powers of the summation index
$n$. To this end we apply
  $$\eqalign{ B_j (x+1/2) &= 2^{1-j} B_j (2x) -B_j
(x) \cr
     &= \sum_{k=0} ^j { j \choose k}(2^{1-k} -1) B_k x^{j-k}   .}
  $$
Noting that $B_{2k+1}=0$ for $k=1,2,...$, and, combining the various
contributions, the final answer for the internal energy on the D+N
quarter sphere (equivalently the `periodic hemisphere') can be cast into
the general form,
  $$
\eqalign{ \overline E_d = \overline E _{d,0} + \sum_{l=0}^{r-1} &I_{l,r}
\sum_{n=1}^\infty
  n^{2l+3} {q^{2n}\over 1-q^{2n}}\cr
  &  + J_r \sum_{n=0}^\infty (2n+1) {q^{2(2n+1)}\over 1-q^{2(2n+1)}}\,,
  }\eql{alldim} $$
where the constants $I_{l,r}$ and $J_r$ are
  $$\eqalign{
I_{l,r}&= \sum_{j=0}^{r-1-l} {c_{l+j}\over 2(l+j+1)} {2+2l+2j
\choose 2j}(1-2^{2j-1})
B_{2j}  ,\cr J_r &= \sum_{k=0}^{r-1} {c_k\over 2(k+1)}(1-2^{2k+2}) B_{2+2k}
. }
  $$
The zero temperature value, $\overline E_{d,0}$, is a generalised
Bernoulli coefficient, [\pref{ChandD}].

This form of the energy, for all dimensions $d$, is then related to the
Fourier expansions of zs $u$ and ns $u$.

In $d=5$ dimensions the answer reads explicitly,
  $$\eqalign{
  \overline E_5 &= \overline E _{5,0} + {1 \over 24} \sum_{n=1} ^\infty
   n^5 {q^{2n} \over 1-q^{2n}} + {1 \over 12} \sum_{n=1} ^\infty
  n^3  {q^{2n} \over 1-q^{2n}} \cr
 & -{1\over 8} \sum_{n=0} ^\infty (2n+1) {q^{2 (2n+1)} \over
1-q^{2(2n+1)}} \,,}
  \eql{en51}$$
which, equivalently from (\peq{en2}), is to be compared with,
  $$ \overline E _5 = \overline E_{5,0} + {1
\over 2^6} (7Q_2 + 5Q_3 +Q_1 -R_1) .
  \eql{en52}$$
Numerical comparison has also been performed as a check of the
calculation.

As mentioned, in $d=3$ dimensions we have to add the $m=0$ contribution
to the result [\peq{alldim}] to get the correct answer. Then one finds,
  $$\overline E_3 = \overline E_{3,0} + {1 \over 2} \sum_{n=1} ^\infty
  n^3 {q^{2n} \over 1-q^{2n}} + {1 \over 2} \sum_{n=0} ^\infty
 (2n+1) {q^{2 (2n+1)} \over 1-q^{2 (2n+1)}}\,,
  \eql{en31}$$
with $\overline E_{3,0}=11/480$.

Alternatively, according to (\peq{en2}),
   $$
   \overline E_{3}=\overline
E_{3,0}+{1\over16}\big(3Q_{1}+3\,Q_{2}+R_1\big)
  \eql{en23}$$
and, again, agreement can be shown since, from (\peq{cue}),
 $$
  {1\over16}\big(3(Q_1+Q_2)+R_1\big)={1\over4}\sum_{n=1} ^\infty
  \big(2n^3+(1-(-1)^n)n\big) {q^{2n} \over 1-q^{2n}}\,.
 $$

For comparison, the energies on the full spheres are
  $$\eqalign{
  \ol E_3\big|_{full}&=\ep_2\cr
 \ol E_5\big|_{full}&={1\over12}\big(\ep_3-\ep_2\big)
 }
  $$
in terms of the partial energies, $\ep_t$, defined in [\pref{DandK1}] as,
  $$
 \ep_t(\xi)=-{B_{2t}\over 4t}+
  \sum_{n=1}^\infty {n^{2t-1}q^{2n}\over1-q^{2n}}\,.
  \eql{inten3}
  $$

As we have mentioned, the computation via the hyperbolic summations is an
unnecessary detour only in that, to evaluate elliptically, it seems that
the degeneracies have to be reintroduced.

\section{\bf 5. Temperature inversion.}

Apart from the final term, the structure of the energy, (\peq{alldim}),
is similar to that on the full sphere and so, for this part, the
conclusions will be the same as in our previous paper, [\pref{DandK1}].
The differences are due to the effect of the polar singularities which
also account for the final term in $E$. Because of this term, the
recursion formula for the $\ep_t$ employed in (\pref{DandK1}) now shows
that any energy can be expressed, as a polynomial in $\ep_2$ and $\ep_3$,
plus a multiple of this last term.

The final term in (\peq{alldim}) also affects the bevaviour under
temperature inversion, $\xi\to1/\xi$. To investigate this, it is
rewritten
  $$
  \sum_{n=0} ^\infty
 (2n+1) {q^{2 (2n+1)} \over 1-q^{2 (2n+1)}}=\sum_{n=0} ^\infty
  n {q^{2n} \over 1-q^{2n}}-\sum_{n=0} ^\infty
  2n {q^{4n} \over 1-q^{4n}}\,,
  $$
and, while the first term on the right has a well defined behaviour under
$\xi\to1/\xi$, described in [\pref{DandK1}], the second apparently does
not, as it corresponds to a different temperature. The conclusion is that
the inversion properties on the sphere do not carry over to its orbifold
factors.

\section{\bf 6. The specific heat and modular covariant derivatives.}

In [\pref{DandK1}] it was shown how the classic elliptic recursions allow
the internal energies on all odd spheres  to be determined in terms of
just two parameters. The same statement holds for the specific heat, as
we demonstrate. From now on we revert to the full sphere.

Before Weierstrass, Eisenstein introduced the double series
  $$
  \caE_n(z)=\sum_{m_1,m_2=-\infty}^\infty{1\over (z+m_1\om_1+m_2\om_2)^n}\,,
  $$
denoting them by $(n,z)$. If $n>2$ they are the higher derivatives of the
Weierstrass $\wp$--function. For the time being we adhere to Eisenstein.
When $n$ equals 1 or 2, the sums are not absolutely convergent and
Eisenstein defines them by a limiting procedure which depends on the
periods, $\om_1$, $\om_2$. We will not expound this here (see Weil
[\pref{Weil}]). It is similar to the later procedures of Glaisher and
Hurwitz.

The basic Eisenstein function, $\caE_n(z)$, has the power series
expansion
  $$
  \caE_n(z)={1\over z^n}+(-1)^n\sum_{t=1}^\infty \comb{2t-1}{n-1}\,\overline
  G_t(\om_1,\om_2)\,z^{2t-n}
  \eql{ps}$$
in terms of the more common Eisenstein series,
  $$
  G_t(\om_1,\om_2)=\sumdasht{{m_1,m_2\atop=-\infty}}{\infty}{1\over
  \big(m_1\om_1+m_2\om_2\big)^{2t}}\,,
  \eql{Eis}$$
by
  $$
  \overline G_t(\om_1,\om_2)=G_t(\om_1,\om_2)\,,\quad t\ge2
  $$
and
  $$
  \overline G_1(\om_1,\om_2)=G_1(\om_1,\om_2)+{\pi i\over\om_1\om_2}\,.
  \eql{gee1}$$
$G_1$ is the quantity that behaves homogeneously under modular
transformations of the periods, \eg [\pref{Rad}].

The $\overline G_t$ are related to the partial energies by,
   $$
 \ep_t(\xi)=(-1)^t{(2t-1)!\over2(2\pi)^{2t}}\, \overline G_t(1,i/\xi)\,,
  \eql{inten4}$$
for all $t$.

By paralleling trigonometric theory, Eisenstein derived a number of basic
identities for the $\caE_n$ including, effectively, the first order
differential equation for $\wp$, some years before Weierstrass. This
equation is the origin of the recursion used in [\pref{DandK1}] and which
is repeated here,
  $$
 \ep_t(\xi)=12{(t-1)(2t-3)\over(2t+1)(t-3)}\sum_{l=2}^{t-2}\comb{2t-4}{2l-2}
 \ep_l(\xi)\,\ep_{t-l}(\xi)\,,\quad t\ge4\,.
  \eql{recurs}$$

However, Eisenstein also derived the important relation for the
derivative with respect to  a period, (see [\pref{Weil}]).
  $$
  {2\pi i\over\om_1}{\pa\caE_1\over\pa\om_2}=\caE_3-\caE_1\,\caE_2\,.
  \eql{ederiv}$$
From this, one can obtain an equation for a `partial specific heat', $\si_t$,
defined by
  $$
  \si_t=\left(\frac{2\pi} \xi\right)^2  \,D\,\ep_t
   $$
in terms of the more convenient quantities,
  $$
  D\,\ep_t=q^2{d\ep_t\over d q^2}
  =-{d\,\ep_t(\xi)\over2\pi d\big({1\over\xi}\big)}={1\over2\pi i}{d\,\ep_t(\tau)
\over d\tau}=-a{d\ep_t(\be)\over d\be}
  \,,\quad t\ge1\,.
  \eql{psh}$$
Our notation is that, for example, $\ep=\ep(\xi)=\ep(\tau)=\ep(\be)$,
depending on whether we wish to use $\xi$, $\tau$ or $\be$ as the
variable.

By choosing $\om_1=1$, $\om_2=i/\xi=\tau=i\be/2\pi a$ and
substituting the power series into (\peq{ederiv}) one easily finds
  $$\eqalign{D\,\ep_1=&-2\ep^2_1+{5\over6}\,\ep_2\cr
  D\,\ep_t=&-4t\,\ep_1\,\ep_t+{2t+3\over2(2t+1)}\,\ep_{t+1}-\sum_{k=2}^{t-1}
  \comb{2t}{2k-1}\ep_k\ep_{t-k+1}\,,\quad t>1\,,}
  \eql{shr}$$
displaying a dependence on $\ep_1$ which is vital for maintaining the
proper inversion behaviour under $\xi\to1/\xi$, which is,
  $$
 \ep_t(1/\xi)=(-1)^t {1\over\xi^{2t}}\,\ep_t(\xi)\,,\quad t\ge2\,.
  \eql{ind}$$
A formal way of describing the situation is to define a `modular
covariant derivative', $\caD$, by
  $$\eqalign{\caD\,\ep_1\equiv&\big(D+2\ep_1\big)\ep_1\cr
  \caD\,\ep_t\equiv&\big(D+4t\ep_1\big)\ep_t\,,\quad t>1\,,}
  \eql{coderiv}$$
so that $\caD\ep_t$ transforms under inversion, $\xi\to1/\xi$, like
$\ep_{t+1}$. Note that, as usual, the covariant derivative depends on the
behaviour of its operand and we shall extend $\caD$ to any modular form
of weight $2t$. (The terminology is that the homogeneous form,
$G_t(\om_1,\om_2)$, is said to have {\it dimension} $-2t$ while the
inhomogeneous, holomorphic form, $G_t(1,\tau)$, has {\it weight} $2t$.)
Although not a modular form, $\ep_1$ can be said to have weight 2. Note
that $\caD\ep_1$ differs from the general structure of $\caD\ep_t$. This
definition is not that used by Lang, [\pref{Lang}], who employs a
covariant derivative, $\pa$, in a discussion of mod $p$ modularity.

The general modular form of weight $k$ is an isobaric polynomial in the
two invariants, $g_2$ and $g_3$, of elliptic function theory. This
follows essentially from the recursion (\peq{recurs}). (Equivalent to
$(g_2,g_3)$ are $(\ep_2,\ep_3)$, $(G_2,G_3)$ and Ramanujan's $(M,N)$,
defined later.) Thus the general form, $\caF$, is written
  $$
  \caF=\sum_{a,b:4a+6b=k}\,\caF_{a,b}\,g_2^a\,g_3^b\,.
  $$

In this setting, the combination in $\caD$ appears in Ogg, [\pref{Ogg2}],
pp.17,18. Ogg uses the modular transformation behaviour of $\ep_1$ to show
that $\caD$ increases the weight by 2 but we could in fact {\it determine}
this behaviour immediately from (\peq{shr}). Hurwitz, [\pref{Hurwitz}] \S4
employs this device.

Higher derivatives can be constructed simply since the usual rules apply.
For example (\peq{shr}) and its derivative read, for {\it all} $t$,
  $$\eqalign{
  \caD\,\ep_t=&{2t+3\over2(2t+1)}\,\ep_{t+1}-\sum_{k=2}^{t-1}
  \comb{2t}{2k-1}\ep_k\ep_{t-k+1}\cr
  \caD^2\ep_t=&{2t+3\over2(2t+1)}\,\caD\ep_{t+1}-2\sum_{k=2}^{t-1}
  \comb{2t}{2k-1}\caD\ep_k\cdot\ep_{t-k+1}\,,}
  \eql{coderivs}$$
and so on. An equivalent set of equations is given by Ramanujan
[\pref{Raman}] pp.165,166, but not using a covariant derivative.

Before drawing any consequences from these equations, some relevant
elliptic facts are interpolated.

The dependence on $\ep_1$ means that the complete elliptic function of
the second kind, $E$, now makes an appearance via,
  $$
  \ep_1=-
  {1\over24}\bigg({2K\over\pi}\bigg)^2\bigg({3E\over K}+k^2-2\bigg)\,.
  \eql{zu1}$$
Zucker, [\pref{zucker2}], from results of Ramanujan, says that $E$, as well
as $K$, can be expressed in terms of algebraic numbers and Gamma functions
at singular moduli.

(\peq{zu1}) can be transformed by use of the standard differential
relation between $K$ and $E$ (\eg Fricke [\pref{Fricke}] I,p.46)
  $$
  2k^2k'^2{dK\over d k^2}=E-k'^2K
  $$
to which we can also add,
  $$
  D={1\over2}\,\big(\gK kk'\big)^2\,{d\over dk^2}\,,\quad \gK={2K\over\pi}\,,
  \eql{diffrel}$$
which ellipticises $D$. In fact, direct manipulation with $q$-series
gives,
  $$
  \ep_1=-{1\over6}D\log\big(kk'\gK^3\big)\,,
  \eql{dede3}$$
which connects with the alternative expression for $\ep_1$,
  $$
  \ep_1=-D\,\log\eta\,,
  \eql{dede1}$$
in terms of Dedekind's $\eta$-function,
  $$
  \eta(\tau)=q^{1/12}\prod_j(1-q^{2j})\,,\quad q=e^{i\pi\tau}=e^{-\pi/\xi}\,.
  \eql{dede2}$$
This includes both the Casimir (zero temperature) contribution and the
statistical mode sum.

We have thus more or less arrived at Jacobi's result
  $$
  \eta^{24}={1\over(2\pi)^{12}}\,\big(g_2^3-27g_3^2\big)={1\over\,2^{8}}\,
  k^4k'^4\gK^{12}\,,
  \eql{etad2}$$
another proof of which occurs shortly.

Returning to (\peq{coderivs}), it is apparent from the standard
recursions, (\peq{recurs}), (\peq{coderivs}), that the ordinary
derivative, $D\,\ep_t$, can be expressed as a polynomial in $\ep_1$,
$\ep_2$ and $\ep_3$ (linear in $\ep_1$, if $t>1$) and the conclusion is
that, at temperatures corresponding to singular moduli, not only the
internal energy but also the specific heat can be expressed in finite
terms in algebraic numbers and gamma functions. Furthermore, the same
statement holds for {\it all} higher derivatives at singular moduli, as
follows by iteration of (\peq{coderivs}) and elimination. Incidentally,
it is, of course, possible to derive (\peq{shr}) in Weierstrassian vein,
and some might prefer this. Such a version is detailed by van der Pol
[\pref{vandp}].

Kaneko and Zagier, [\pref{KandZ}], refer to the algebra generated by
$\ep_1$, $\ep_2$ and $\ep_3$ as the algebra of {\it quasi--modular
forms.}

For the circle, the three-sphere and the five-sphere we find for
the total specific heats for conformal scalars, \mgn{DO THIS
PROPERLY?? Are we interested?}
  $$\eqalign{
  \si^{(1)}(\xi)&=\left( \frac {2\pi} \xi \right)^2
  \left[ -4\ep_1^2+{5\over3}\,\ep_2\right]\cr
  \si^{(3)}(\xi)&=\left( \frac {2\pi} \xi \right)^2
  \left[ -8\ep_1\ep_2+{7\over10}\,\ep_3\right]\cr
  \si^{(5)}(\xi)&=\left( \frac {2\pi} \xi \right)^2
  \left[ -\ep_1 \ep_3 + \frac {100} {21} \ep_2^2 + \frac 2 3 \ep_1 \ep_2 -
  \frac 7 {120} \ep_3\right]\,.}
  \eql{spht}
  $$

As a particular numerical case, at the lemniscate point, $\xi=1$, (see
[\pref{DandK1}]), one finds
  $$
  \si^{(1)}(1)={1\over1152\pi^4}\big(\Ga^8(1/4)+24\pi^2\Ga^4(1/4)-288\pi^4\big)
  \approx 0.380810377\,.
  $$
Other examples we leave to the reader.

Although equation (\peq{shr}) provides a systematic method, another means
of finding the derivative in (\peq{psh}) is to apply it to the explicit
forms of the partial energies, $\ep_t$, in terms of $k$ and $K$. This is
done by Glaisher [\pref{Glaisher}], \S\S78,79, who gives the necessary
rules. One can easily see that the general conclusion will be the same.
The structure of $\ep_t$ is a polynomial in $k^2$ multiplied by a power
of $K$. The ellipticised form of $D$, (\peq{diffrel}), shows that $\ep_1$
will appear through the action of $d/d k^2$ on the $K$ factor.

Yet another method is to write the Eisenstein function, $\caE_n$, in
terms of the $\th_1$--function, and then use the differential (heat)
equation that this satisfies, or use standard theta--function identities.
In modern parlance this is an approach via {\it Jacobi forms}, which are
functions of {\it two} variables. We remark that instead of the
$\th$--functions one can use the Jacobi elliptic functions, sn $u$, \etc
The physical significance of the $u$ variable could be the radial
separation of source and field point on the sphere (\eg Allen {\it et al}
[\pref{Allen}]).

We now make some further remarks on the recursions and their history.
Because of the recursions, it is sufficient to give the derivatives of
just $\ep_1$, $\ep_2$ and $\ep_3$, and, in order to express these with
reasonable factors, it is convenient, and conventional, to renormalise
them. We will use Ramanujan's $L$, $M$ and $N$ notation where
$L=-24\ep_1$, $M=240\ep_2$ and $N=-504\ep_3$. Then from Eisenstein's
equation, (\peq{shr}), one rapidly finds the relations often employed by
Ramanujan,
  $$\eqalign{
  D\,L=&{1\over12}(L^2-M)\,,\quad{\rm or}\quad
  \caD\,L=-{1\over12}M\cr
  D\,M=&{1\over3}(LM-N)\,,\quad{\rm or}\quad \caD\,M=-{1\over3}N\cr
  D\,N=&{1\over2}(LN-M^2)\,,\quad{\rm or}\quad \caD\,N=-{1\over2}M^2\,.}
  \eql{3derivs}$$

An important result now follows easily on noting that from (\peq{3derivs})
  $$
  D\log(M^3-N^2)=L\,,
  \eql{reln1}$$
and so, from (\peq{dede1}),
  $$
  D\,\log(M^3-N^2)=24\,D\,\log\eta\,,
  $$
whence, again, Jacobi's relation between the $\eta$--function and the
discriminant, $\De=g_2^3-27\,g_3^2$,
  $$
  \eta^{24}={1\over1728}\,(M^3-N^2)={1\over(2\pi)^{12}}\,\De\,,
  \eql{etad}$$
the 1728 being the coefficient of $q^2$ in $M^3-N^2$.\mgn{Fricke p.274. His
omegas are ours except $1\leftrightarrow2$.}

The Weierstrassian way of deriving this relation is somewhat more involved,
\eg [\pref{Cox}], [\pref{Ogg}]. Even Weil's reconstruction of Eisenstein's
approach mimics this $\wp$ method, [\pref{Weil}], p.33. The modular form
proof, [\pref{Ogg,Rankin, Schoenberg, Serre}], is quite different and
depends on the uniqueness of cusp forms of weight $12$.

Equation (\peq{reln1}) is equivalent to the statement that $\De$ is
covariant constant, (put $2t=12$ in (\peq{coderiv})),
  $$
  \caD\,\De=(D-L)\De=0\,,
  \eql{covconst}$$
which highlights the special connecting role played by $L$. This
condition also follows, as noted by Tuite, [\pref{Tuite}], from the fact
that there are no (elliptic) cusp forms of weight 14 and so works the
argument in reverse.

Equation (\peq{covconst}) is a disguised version of the statistical
mechanical relation,
  $
  E=-{\pa \log Z/\pa\be}\,,
  $
on the circle.

Incidentally, the use of $D$ and $\caD$ allows one to give a more general
and systematic treatment of Ramanujan's Notebook  Entry 45 (p.352), see
[\pref{Berndt3}] part V, p.484.  Also, as a side comment, we note that
equations (\peq{3derivs}) and (\peq{reln1}) allow  derivations of the
nonlinear differential equations satisfied by $M$, $N$ and $\De$, \eg
[\pref{Resnikoff,vandp}], relatively easily.

The relations (\peq{recurs}) and (\peq{shr}) were first obtained by
Eisenstein\mgn{CHECK}. Later derivations have been given by Ramanujan,
[\pref{Raman}], as mentioned, van der Pol [\pref{vandp}], Rankin,
[\pref{Rankin2}], and Skoruppa [\pref{Skoruppa}]. Rankin and Skoruppa use a
number theoretic approach, aspects of which are reminiscent of Glaisher's
work.
\section{\bf 7. Elliptic formulation of the free energy on the circle.}

Equation (\peq{dede1}) allows one to obtain an expression for a `partial
free energy', $f_1$, defined by,
  $$
  \ep_1={\pa\over\pa{1\over\xi}}\,\bigg({1\over\xi}f_1\bigg)
  $$
\ie
  $$
   f_1={\xi\over2\pi}\log\eta\,,
  \eql{parfreen}$$

On the circle the total (scaled) free energy is the well known equation,
  $$
  \overline F_1=2f_1-{\xi\over2\pi}\log\xi={\xi\over2\pi}\log(\eta^2/\xi)\,.
  \eql{tfreen}$$
There is a degeneracy factor of 2, and the $\log\xi$ term is the
zero mode contribution.

Relation (\peq{etad}) now enables one to give an elliptic interpretation to
the partial free energy, (\peq{parfreen}), by
  $$
  f_1={\xi\over48\pi}\,\log (\De/c)\,,\quad c=(2\pi)^{12}\,.
  \eql{parfreen2}$$

Thus the free energy on the circle can be thought of as a function, but not
an algebraic one, of the elliptic modulus, $k$, remembering that
$\xi=K/K'$. See the explicit form in (\peq{dede3}), (\peq{etad2}).

The choice of constants, in (\peq{parfreen}), is governed by the desire
to satisfy Nernst's theorem \ie to make the entropy vanish at absolute
zero, $\xi=0$. A partial entropy on the circle can be defined by $ s_1=
(f_1-e_1)/\xi=- d{} f_1/d\xi$ which is again elliptic. It should be said
that, when we define these `partial quantities', we are leaving the zero
mode contribution in (\peq{tfreen}) aside as it causes problems with
Nernst's theorem which have to be addressed separately.

\section{\bf 8. Higher sphere thermodynamics.}
The question is now whether the higher sphere statistical mechanics can
be `ellipticised' in the same way as on the circle. The internal energy
has already been treated in [\pref{DandK1}] but the problem is the free
energy, and thence the entropy, which demands an effective integration.

It must be remarked however that Zucker, [\pref{zucker}], following
Selberg and Chowla, uses (\peq{etad2}) to {\it derive} closed form values
for $K$ at singular moduli from computed values of $\eta$ via the Epstein
\zf. If this were the only method, then the elliptic character of the
circle free energy would be secondary. However there are other ways of
finding the singular $K$ which do not involve Kronecker's limit formula.
We can still therefore maintain the attitude of our earlier work,
[\pref{DandK1}], and regard the question of the  possible ellipticisation
of the free energy on {\it all} odd spheres as motivation for further
investigation. This will be undertaken in a further more technical
communication.

\section{\bf 8. Summary.}

It has been shown that the temperature inversion properties of the
internal energy on odd spheres does not carry through to their orbifold
factors. The specific case of the quarter sphere is considered in detail.
During the analysis, hyperbolic summations arise that have not so far
been encountered.

It is also proved that the specific heat on any odd (full) sphere can be
expressed in terms of just the three partial energies, $\ep_1,\,\ep_2$
and $\ep_3$, the novelty being the appearance of $\ep_1$, related to the
discriminant The derivation makes use of the notion of modular covariant
derivative with $\ep_1$ as a `connection'. We note, incidentally, that
the discriminant is covariant constant.

Finally we have demonstrated that the free energy on the circle can be
ellipticised but have left the corresponding statement for all (odd)
spheres open and subject to later analysis.

\section{\bf References.}
\begin{putreferences}
  \ref{Tuite}{Tuite,M.P. {\it Genus two meromorphic conformal field
  theory}. \break ArXiv:math /9910136v2.}
  \ref{KandZ}{Kaneko,M. and Zagier,D. {\it A generalized Jacobi theta
  function and quasimodular forms.} The Moduli Space of Curves (Texel
  Island, 1994) (Birkhauser, Boston, 1995).}
  \ref{DandA}{Dowker,J.S. and Apps, J.S. \cqg{12}{1995}{1363}.}
  \ref{Weil}{Weil,A., {\it Elliptic functions according to Eisenstein and
  Kronecker}, (Springer, Berlin, 1976).}
  \ref{Ling}{Ling,C-B.{\it SIAM J.Math.Anal.} {\bf5} (1974) 551.}
  \ref{Ling2}{Ling,C-B. {\it SIAM J.Math.Anal.} {\bf6} (1975) 129;
  {\it J.Math.Anal.Appl.} {\bf 13} (1988) 451.}
 \ref{BMO}{Brevik,I., Milton,K.A. and Odintsov, S.D. {\it Entropy bounds in
$R\times S^3$ geometries}. hep-th/0202048.}
 \ref{KandL}{Kutasov,D. and Larsen,F. {\it JHEP} 0101 (2001) 1.}
 \ref{KPS}{Klemm,D., Petkou,A.C. and Siopsis {\it Entropy
bounds, monoticity properties and scaling in CFT's}. hep-th/0101076.}
\ref{DandC}{Dowker,J.S. and Critchley,R. \prD{15}{1976}{1484}.}
\ref{AandD}{Al'taie, M.B. and Dowker, J.S. \prD{18}{1978}{3557}.}
\ref{Dow1}{Dowker,J.S. \prD{37}{1988}{558}.}
\ref{Dow3}{Dowker,J.S. \prD{28}{1983}{3013}.}
\ref{DandK}{Dowker,J.S. and Kennedy,G. \jpa{}{1978}{}.}
\ref{Dow2}{Dowker,J.S. \cqg{1}{1984}{359}.}
\ref{DandKi}{Dowker,J.S. and Kirsten, K.{\it Comm. in Anal. and Geom.
}{\bf7}(1999) 641.}
\ref{DandKe}{Dowker,J.S. and Kennedy,G. \jpa{11}{1978}{895}.}
\ref{Gibbons}{Gibbons,G.W. \pl{60A}{1977}{385}.}
\ref{Cardy}{Cardy,J.L. \np{366}{1991}{403}.}
\ref{ChandD}{Chang,P. and Dowker,J.S. \np{395}{1993}{407}.}
\ref{DandC2}{Dowker,J.S. and Critchley,R. \prD{13}{1976}{224}.}
\ref{Camporesi}{Camporesi,R. \prp{196}{1990}{1}.}
\ref{BandM}{Brown,L.S. and Maclay,G.J. \pr{184}{1969}{1272}.}
\ref{CandD}{Candelas,P. and Dowker,J.S. \prD{19}{1979}{2902}.}
\ref{Unwin1}{Unwin,S.D. Thesis. University of Manchester. 1979.}
\ref{Unwin2}{Unwin,S.D. \jpa{13}{1980}{313}.}
\ref{DandB}{Dowker,J.S. and Banach,R. \jpa{11}{1979}{}.}
\ref{Obhukov}{Obhukov,Yu.N. \pl{109B}{1982}{195}.}
\ref{Kennedy}{Kennedy,G. \prD{23}{1981}{2884}.}
\ref{CandT}{Copeland,E. and Toms,D.J. \np {255}{1985}{201}.}
\ref{ELV}{Elizalde,E., Lygren, M. and Vassilevich, D.V. \jmp {37}{1996}{3105}.}
\ref{Malurkar}{Malurkar,S.L. {\it J.Ind.Math.Soc} {\bf16} (1925/26) 130.}
\ref{Glaisher}{Glaisher,J.W.L. {\it Messenger of Math.} {\bf18} (1889) 1.}
\ref{Anderson}{Anderson,A. \prD{37}{1988}{536}.}
 \ref{CandA}{Cappelli,A. and D'Appollonio,\pl{487B}{2000}{87}.}
 \ref{Wot}{Wotzasek,C. \jpa{23}{1990}{1627}.}
 \ref{RandT}{Ravndal,F. and Tollesen,D. \prD{40}{1989}{4191}.}
 \ref{SandT}{Santos,F.C. and Tort,A.C. \pl{482B}{2000}{323}.}
 \ref{FandO}{Fukushima,K. and Ohta,K. {\it Physica} {\bf A299} (2001) 455.}
 \ref{GandP}{Gibbons,G.W. and Perry,M. \prs{358}{1978}{467}.}
 \ref{Dow4}{Dowker,J.S. {\it Zero modes, entropy bounds and partition
functions.} hep-th\break /0203026.}
  \ref{Rad}{Rademacher,H. {\it Topics in analytic number theory,}
(Springer- Verlag,  Berlin, 1973).}
  \ref{Halphen}{Halphen,G.-H. {\it Trait\'e des Fonctions Elliptiques}, Vol 1,
Gauthier-Villars, Paris, 1886.}
  \ref{CandW}{Cahn,R.S. and Wolf,J.A. {\it Comm.Mat.Helv.} {\bf 51} (1976) 1.}
  \ref{Berndt}{Berndt,B.C. \rmjm{7}{1977}{147}.}
  \ref{Hurwitz}{Hurwitz,A. \ma{18}{1881}{528}.}
  \ref{Hurwitz2}{Hurwitz,A. {\it Mathematische Werke} Vol.I. Basel,
  Birkhauser, 1932.}
  \ref{Berndt2}{Berndt,B.C. \jram{303/304}{1978}{332}.}
  \ref{RandA}{Rao,M.B. and Ayyar,M.V. \jims{15}{1923/24}{150}.}
  \ref{Hardy}{Hardy,G.H. \jlms{3}{1928}{238}.}
  \ref{TandM}{Tannery,J. and Molk,J. {\it Fonctions Elliptiques},
   Gauthier-Villars, Paris, 1893--1902.}
  \ref{schwarz}{Schwarz,H.-A. {\it Formeln und Lehrs\"atzen zum Gebrauche..},
  Springer 1893.(The first edition was 1885.) The French translation by
Henri Pad\'e is
{\it Formules et Propositions pour L'Emploi...}, Gauthier-Villars, Paris,
1894}
  \ref{Hancock}{Hancock,H. {\it Theory of elliptic functions}, Vol I.
   (Wiley, New York 1910).}
  \ref{watson}{Watson,G.N. \jlms{3}{1928}{216}.}
  \ref{MandO}{Magnus,W. and Oberhettinger,F. {\it Formeln und S\"atze},
  Springer-Verlag, Berlin 1948.}
  \ref{Klein}{Klein,F. {\it }.}
  \ref{AandL}{Appell,P. and Lacour,E. {\it Fonctions Elliptiques},
  Gauthier-Villars,
  Paris, 1897.}
  \ref{HandC}{Hurwitz,A. and Courant,C. {\it Allgemeine Funktionentheorie},
  Springer,
  Berlin, 1922.}
  \ref{WandW}{Whittaker,E.T. and Watson,G.N. {\it Modern analysis},
  Cambridge 1927.}
  \ref{SandC}{Selberg,A. and Chowla,S. \jram{227}{1967}{86}. }
  \ref{zucker}{Zucker,I.J. {\it Math.Proc.Camb.Phil.Soc} {\bf 82 }(1977) 111.}
  \ref{glasser}{Glasser,M.L. {\it Maths.of Comp.} {\bf 25} (1971) 533.}
  \ref{GandW}{Glasser, M.L. and Wood,V.E. {\it Maths of Comp.} {\bf 25} (1971)
  535.}
  \ref{greenhill}{Greenhill,A,G. {\it The Applications of Elliptic
  Functions}, MacMillan, London, 1892.}
  \ref{Weierstrass}{Weierstrass,K. {\it J.f.Mathematik (Crelle)}
{\bf 52} (1856) 346.}
  \ref{Weierstrass2}{Weierstrass,K. {\it Mathematische Werke} Vol.I,p.1,
  Mayer u. M\"uller, Berlin, 1894.}
  \ref{Fricke}{Fricke,R. {\it Die Elliptische Funktionen und Ihre Anwendungen},
    (Teubner, Leipzig. 1915, 1922).}
  \ref{Konig}{K\"onigsberger,L. {\it Vorlesungen \"uber die Theorie der
 Elliptischen Funktionen},  \break Teubner, Leipzig, 1874.}
  \ref{Milne}{Milne,S.C. {\it The Ramanujan Journal} {\bf 6} (2002) 7-149.}
  \ref{Schlomilch}{Schl\"omilch,O. {\it Ber. Verh. K. Sachs. Gesell. Wiss.
  Leipzig}  {\bf 29} (1877) 101-105; {\it Compendium der h\"oheren Analysis},
  Bd.II, 3rd Edn, Vieweg, Brunswick, 1878.}
  \ref{BandB}{Briot,C. and Bouquet,C. {\it Th\`eorie des Fonctions Elliptiques},
  Gauthier-Villars, Paris, 1875.}
  \ref{Dumont}{Dumont,D. \aim {41}{1981}{1}.}
  \ref{Andre}{Andr\'e,D. {\it Ann.\'Ecole Normale Superior} {\bf 6} (1877) 265;
  {\it J.Math.Pures et Appl.} {\bf 5} (1878) 31.}
  \ref{Raman}{Ramanujan,S. {\it Trans.Camb.Phil.Soc.} {\bf 22} (1916) 159;
 {\it Collected Papers}, (C.U.P., Cambridge, 1927).}
  \ref{Weber}{Weber,H.M. {\it Lehrbuch der Algebra} Bd.III, Vieweg,
  Brunswick 190  3.}
  \ref{Weber2}{Weber,H.M. {\it Elliptische Funktionen und algebraische Zahlen},
  Vieweg, Brunswick 1891.}
  \ref{ZandR}{Zucker,I.J. and Robertson,M.M.
  {\it Math.Proc.Camb.Phil.Soc} {\bf 95 }(1984) 5.}
  \ref{JandZ1}{Joyce,G.S. and Zucker,I.J.
  {\it Math.Proc.Camb.Phil.Soc} {\bf 109 }(1991) 257.}
  \ref{JandZ2}{Zucker,I.J. and Joyce.G.S.
  {\it Math.Proc.Camb.Phil.Soc} {\bf 131 }(2001) 309.}
  \ref{zucker2}{Zucker,I.J. {\it SIAM J.Math.Anal.} {\bf 10} (1979) 192,}
  \ref{BandZ}{Borwein,J.M. and Zucker,I.J. {\it IMA J.Math.Anal.} {\bf 12}
  (1992) 519.}
  \ref{Cox}{Cox,D.A. {\it Primes of the form $x^2+n\,y^2$}, (Wiley, New York,
  1989).}
  \ref{BandCh}{Berndt,B.C. and Chan,H.H. {\it Mathematika} {\bf42} (1995) 278.}
  \ref{EandT}{Elizalde,R. and Tort.hep-th/}
  \ref{KandS}{Kiyek,K. and Schmidt,H. {\it Arch.Math.} {\bf 18} (1967) 438.}
  \ref{Oshima}{Oshima,K. \prD{46}{1992}{4765}.}
  \ref{greenhill2}{Greenhill,A.G. \plms{19} {1888} {301}.}
  \ref{Russell}{Russell,R. \plms{19} {1888} {91}.}
  \ref{BandB}{Borwein,J.M. and Borwein,P.B. {\it Pi and the AGM}, Wiley,
  New York, 1998.}
  \ref{Resnikoff}{Resnikoff,H.L. \tams{124}{1966}{334}.}
  \ref{vandp}{Van der Pol, B. {\it Indag.Math.} {\bf18} (1951) 261,272.}
  \ref{Rankin}{Rankin,R.A. {\it Modular forms} (CUP, Cambridge,1977).}
  \ref{Rankin2}{Rankin,R.A. {\it Proc. Roy.Soc. Edin.} {\bf76 A} (1976) 107.}
  \ref{Skoruppa}{Skoruppa,N-P. {\it J.of Number Th.} {\bf43} (1993) 68 .}
  \ref{Down}{Dowker.J.S. \np {104}{2002}{153}.}
  \ref{Eichler}{Eichler,M. \mz {67}{1957}{267}.}
  \ref{Zagier}{Zagier,D. \invm{104}{1991}{449}.}
  \ref{Lang}{Lang,S. {\it Modular Forms}, (Springer, Berlin, 1976).}
  \ref{Kosh}{Koshliakov,N.S. {\it Mess.of Math.} {\bf 58} (1928) 1.}
  \ref{BandH}{Bodendiek, R. and Halbritter,U. \amsh{38}{1972}{147}.}
  \ref{Smart}{Smart,L.R., \pgma{14}{1973}{1}.}
  \ref{Grosswald}{Grosswald,E. {\it Acta. Arith.} {\bf 21} (1972) 25.}
  \ref{Kata}{Katayama,K. {\it Acta Arith.} {\bf 22} (1973) 149.}
  \ref{Ogg}{Ogg,A. {\it Modular forms and Dirichlet series} (Benjamin, New York,
   1969).}
  \ref{Bol}{Bol,G. \amsh{16}{1949}{1}.}
  \ref{Epstein}{Epstein,P. \ma{}{}{}}
  \ref{Petersson}{Petersson.}
  \ref{Serre}{Serre,J-P. {\it A Course in Arithmetic}, (Springer, New York,
  1973).}
  \ref{Schoenberg}{Schoenberg,B., {\it Elliptic Modular Functions},
  (Springer, Berlin, 1974).}
  \ref{Apostol}{Apostol,T.M. \dmj {17}{1950}{147}.}
  \ref{Ogg2}{Ogg,A. {\it Lecture Notes in Math.} {\bf 320} (1973) 1.}
  \ref{Knopp}{Knopp,M.I. \dmj {45}{1978}{47}.}
  \ref{Knopp2}{Knopp,M.I. \invm {}{1994}{361}.}
  \ref{LandZ}{Lewis,J. and Zagier,D. \aom{153}{2001}{191}.}
  \ref{DandK1}{Dowker,J.S. and Kirsten,K. \np{638}{2002}{405}.}
  \ref{HandK}{Husseini and Knopp.}
  \ref{Kober}{Kober,H. \mz{39}{1934-5}{609}.}
  \ref{HandL}{Hardy,G.H. and Littlewood, \am{41}{1917}{119}.}
  \ref{Watson}{Watson,G.N. \qjm{2}{1931}{300}.}
  \ref{SandC2}{Chowla,S. and Selberg,A. {\it Proc.Nat.Acad.} {\bf 35}
  (1949) 371.}
  \ref{Landau}{Landau, E. {\it Lehre von der Verteilung der Primzahlen},
  (Teubner, Leipzig, 1909).}
  \ref{Berndt4}{Berndt,B.C. \tams {146}{1969}{323}.}
  \ref{Berndt3}{Berndt,B.C. {\it Ramanujan's Notebooks} edited by B.C.Berndt, pt.V. (Springer, New York,
   1998).}
  \ref{Bochner}{Bochner,S. \aom{53}{1951}{332}.}
  \ref{Weil2}{Weil,A.\ma{168}{1967}{}.}
  \ref{CandN}{Chandrasekharan,K. and Narasimhan,R. \aom{74}{1961}{1}.}
  \ref{Rankin3}{Rankin,R.A. {} {} ().}
  \ref{Berndt6}{Berndt,B.C. {\it Trans.Edin.Math.Soc}.}
  \ref{Elizalde}{Elizalde,E. {\it Ten Physical Applications of Spectral
  Zeta Function Theory}, \break (Springer, Berlin, 1995).}
  \ref{Allen}{Allen,B., Folacci,A. and Gibbons,G.W. \pl{189}{1987}{304}.}
  \ref{Krazer}{Krazer}
  \ref{Elizalde3}{Elizalde,E. {\it J.Comp.and Appl. Math.} {\bf 118} (2000) 125.}
  \ref{Elizalde2}{Elizalde,E., Odintsov.S.D, Romeo, A. and Bytsenko, A.A and
  Zerbini,S.
  {\it Zeta function regularisation}, (World Scientific, Singapore, 1994).}
  \ref{Eisenstein}{Eisenstein}
  \ref{Hecke}{Hecke,E. \ma{112}{1936}{664}.}
  \ref{Terras}{Terras,A. {\it Harmonic analysis on Symmetric Spaces} (Springer,
  New York, 1985).}
  \ref{BandG}{Bateman,P.T. and Grosswald,E. {\it Acta Arith.} {\bf 9} (1964) 365.}
  \ref{Deuring}{Deuring,M. \aom{38}{1937}{585}.}
  \ref{Guinand}{Guinand.}
  \ref{Guinand2}{Guinand.}
  \ref{Minak}{Minakshisundaram.}
  \ref{Mordell}{Mordell,J. \prs{}{}{}.}
  \ref{GandZ}{Glasser,M.L. and Zucker, {}.}
  \ref{Landau2}{Landau,E. \jram{}{1903}{64}.}
  \ref{Kirsten1}{Kirsten,K. \jmp{35}{1994}{459}.}
  \ref{Sommer}{Sommer,J. {\it Vorlesungen \"uber Zahlentheorie} (1907,Teubner,Leipzig).
  French edition 1913 .}
  \ref{Reid}{Reid,L.W. {\it Theory of Algebraic Numbers}, (1910,MacMillan,New York).}
\end{putreferences}
\bye